\documentclass{aastex}
\usepackage{emulateapj5}
\usepackage{apjfonts}
\usepackage{epsf}
\usepackage{times}

\usepackage{graphics}

\newcommand{\be}{\begin{equation}}
\newcommand{\ee}{\end{equation}}
\newcommand{\bd}{\begin{displaymath}}
\newcommand{\ed}{\end{displaymath}}

\slugcomment{Received ; accepted 2004 June 9}

\shorttitle{Are the jets accelerated from the disk coronas?}
\shortauthors{Cao X}

\begin{document}

\title{Are the jets accelerated from the disk coronas in some active galactic
nuclei?}

\author{Xinwu Cao}
\affil{Shanghai Astronomical Observatory, Chinese Academy of
Sciences,
80 Nandan Road, Shanghai, 200030, China\\
Email: cxw@center.shao.ac.cn}

\begin{abstract}

We use a sample of radio-loud active galactic nuclei (AGNs) with
estimated central black hole masses to explore their jet formation
mechanisms. The jet power of AGNs is estimated from their extended
radio luminosity. It is found that the jets in several AGNs of
this sample are too powerful to be extracted from the standard
thin accretion disks or rapidly spinning black holes surrounded by
standard thin disks. If the advection dominated accretion flows
(ADAFs) are present in these AGNs, their bright optical continuum
luminosity cannot be produced by pure-ADAFs due to their low
accretion rates and low radiation efficiency, unless the ADAFs
transit to standard thin disks at some radii $R_{\rm tr}$. If this
is the case, we find that the dimensionless accretion rates
$\dot{m}=\dot{M}/\dot{M}_{\rm Edd}$ as high as $\ge 0.05$ and
transition from ADAFs to standard thin disks at rather small radii
around $\sim 20GM_{\rm bh}/c^2$ are required to explain their
bright optical continuum emission. We propose that the disk-corona
structure is present at least in some AGNs in this sample. The
plasmas in the corona are very hot, and the pressure scale-height
of the corona $H_{\rm c}\sim R$. Powerful jets with $Q_{\rm
jet}\sim L_{\rm bol}$ (bolometric luminosity) can form by the
large-scale magnetic fields created by dynamo processes in the
disk corona of some AGNs. The maximal jet power extractable from
the corona $Q_{\rm jet}^{\rm max}\le 0.6L_{\rm c}$ ($L_{\rm c}$ is
the corona luminosity) is expected by this jet formation scenario.
The statistic results on the sample of AGNs are consistent with
the predictions of this scenario. Finally, the possibility that
the jet is driven from a super-Keplerian rotating hot layer
located between the corona and the cold disk is discussed. We find
that, in principle, this layer can also produce a powerful jet
with $Q_{\rm jet}\sim L_{\rm bol}$.

\end{abstract}


\section{Introduction}

Relativistic jets have been observed in many radio-loud AGNs and
are believed to be formed very close to the black holes. In
currently most favored models of the formation of the jet, the
power is generated through accretion and then extracted from the
disk/black hole rotational energy and converted into the kinetic
power of the jet, namely, the Blandford-Payne and Blandford-Znajek
mechanisms \citep{bp82,bz77}. Both these two jet formation
mechanisms predict a link between the accretion disk and jet. The
disk-jet connection has been investigated by many authors using
observational data in different ways (e.g., Rawlings \& Saunders
1991; Falcke \& Biermann 1995; Xu, Livio, \& Baum 1999; Cao \&
Jiang 1999; 2001), which indicates an intrinsic link between
accretion disks and jets. However, this cannot rule out the
Blandford-Znajek jet formation mechanism.  The magnetic fields are
maintained by the currents in the accretion disk surrounding the
rapidly spinning black hole, so the power extracted from a rapidly
spinning black hole by the Blandford-Znajek mechanism depends on
the properties of the disk near the black hole as well.

It's still unclear which mechanism is responsible for jet
formation in AGNs. The relative importance of these two jet
formation mechanisms is explored by different authors (e.g., Ghosh
\& Abramowicz 1997; Livio, Ogilivie, \& Pringle 1999, hereafter
L99; Cao 2002b). \citet{ga97} doubted the importance of the
Blandford-Znajek process. For a black hole of given mass and
angular momentum, the strength of the Blandford-Znajek process
depends crucially on the strength of the poloidal field threading
the horizon of the hole. The magnetic field threading a hole
should be maintained by the currents situated in the inner region
of the surrounding accretion disk. They argued that the strength
of the field threading a black hole has been overestimated.
\citet{l99} re-investigated the problem and pointed out that even
the calculations of \citet{ga97} have overestimated the power of
the Blandford-Znajek process, since they have overestimated the
strength of the large-scale field threading the inner region of an
accretion disk, and then the efficiency of the Blandford-Znajek
process. The length scale of the fields created by dynamo
processes is of the order of the disk thickness $\sim H$. The
large-scale field can be produced from the small-scale field
created by dynamo processes as $B(\lambda)\propto \lambda^{-1}$
for the idealized case, where $\lambda$ is the length scale of the
field \citep{tp96,r98}. So, the large-scale field is very weak, if
the field is created in the  thin accretion disks. L99 estimated
the maximal jet power extracted from an accretion disk on the
assumption that the toroidal field component is of the same order
of the poloidal field component at the disk surface. They argued
that the maximal jet power extracted from an accretion disk (the
Blandford-Payne mechanism) dominates over the maximal power
extracted by the Blandford-Znajek process (L99). For the ADAF
cases, the disk thickness $H\sim R$ and the jets can be driven by
the large-scale magnetic fields created by dynamo processes (e.g.,
Armitage \& Natarajan 1999). The maximal jet power extracted from
rotating black holes and ADAFs was calculated by \citet{an99} and
\citet{m01}. Instead of the ADAF model, \citet{mf02} proposed that
the jets in low-luminosity AGNs may be magnetically (or thermally)
driven from the coronas above the geometrically thin, optically
thin disks accreting at low rates. { The gas in the corona is
almost virilized, and the thickness of the corona is $\sim R$. The
large-scale  magnetic fields created by dynamo processes in the
corona are significantly stronger than the thin disk due to the
fact of the corona thickness being much larger than the cold thin
disk, and  the disk corona may therefore power a stronger jet than
the thin disk. In principle, the maximal jet power can be
extracted for different jet formation mechanisms can be calculated
if the central black hole mass and accretion rate are known.}

{There are different approaches proposed to measure the black hole
masses in AGNs. The tight correlation between central black hole
mass $M_{\rm bh}$ and stellar dispersion velocity $\sigma$ of the
host galaxy is used to estimate the black hole mass for the
sources with measured stellar dispersion velocity (e.g., Ferrarese
\& Merritt 2000; Gebhardt et al. 2000). The central black hole
mass can also be estimated from its host galaxy luminosity by
using the correlation between black hole mass $M_{\rm bh}$ and
host galaxy luminosity $M_{R}$ at $R$-band (e.g., McLure \& Dunlop
2002). However, there are only a small fraction of AGNs with
measured stellar dispersion velocity or host galaxy luminosity,
which prevents us from using these approaches to estimate the
masses of black holes in most AGNs. \citet{k00} measured the sizes
of broad-line regions (BLRs) in several tens Seyfert 1 galaxies
and quasars from the time delay between their variabilities in
optical continuum emission and broad-line emission using the
reverberation mapping method \citep{p93}. The central black hole
masses of these AGNs have been estimated from their broad-line
widths on the assumption that the motion of the clouds in BLR are
virilized \citep{k00}. They found a correlation between the
measured BLR size and optical continuum luminosity for their
radio-quiet AGN sample. As only a small fraction of AGNs have
measured BLR sizes, this empirical relation between $R_{\rm BLR}$
and optical continuum luminosity $\lambda L_{\lambda}$ are
employed to estimate the masses of black holes in AGNs from their
broad-line widths (e.g., Laor 2000; McLure \& Dunlop 2001; Gu,
Cao, \& Jiang 2001). \citet{md01} compared the virilized H$\beta$
black hole mass estimates with those estimated from $M_{\rm
bh}-\sigma$ relation for a sample of radio-loud AGNs. They found
that the virilized H$\beta$ black hole masses are systematically
lower than that derived from the stellar velocity dispersion, and
they concluded that the disk-like BLR geometry may be in these
sources. This is in general consistent with the anti-correlation
between broad H$\beta$ line width and $R_{\rm c}$ (the ratio of
the strengths of the radio core to the lobe) for a flat-spectrum
radio-loud AGNs found by \citet{wb86}. The inclination angles of
the jets in flat-spectrum AGNs are believed to be small with
respect to the line of sight, and their optical continuum emission
may probably beamed (see Urry \& Padovani 1995, and references
therein). The black hole mass estimate for flat-spectrum
radio-loud AGNs using the broad-line widths and optical continuum
luminosities may be lower than their real values, which implies
that black hole mass estimated from the broad-line width may be
affected by the disk-like BLR orientation \citep{md01}. }

In this work, we use a sample of AGNs to explore the jet formation
in  these AGNs. The black hole masses of the sources in this
sample are estimated from their optical continuum luminosities and
the widths of broad-line H$\beta$.

The sample we use in this investigation is radio-selected, which
includes some sources with very strong jets. In this work, we will
focus on how the powerful jets are formed in the regions near
black holes in these sources.

\section{Black hole masses and accretion rates}

\citet{k00} derived an empirical relation between the BLR size and
optical continuum luminosity \be R_{\rm
BLR}=(32.9^{+2.0}_{-1.9})\left[ {\frac {\lambda L_{\lambda}(5100
{\AA})}{10^{44}{\rm ergs~ s^{-1}}}}\right]^{0.700\pm0.033} {\rm
lt-days}, \label{rblr} \ee for a sample of Seyfert 1 galaxies and
quasars, in which the sizes of BLRs are measured with the
reverberation mapping method \citep{p93}. The central black hole
masses $M_{\rm bh}$ can be estimated from the velocities $V_{\rm
BLR}$ of the clouds in the BLRs \be M_{\rm bh}={\frac {V_{\rm
BLR}^2R_{\rm BLR}}{G}}, \label{mbh}\ee where the motions of the
clouds are assumed to be virilized and isotropic.  The velocities
of the clouds in BLRs $V_{\rm BLR}$ are derived from the width of
the broad emission lines.  For most AGNs, the BLR sizes have not
been measured by the reverberation mapping method, the empirical
relation (\ref{rblr}) is instead used to estimate the BLR sizes.
The central black hole masses of AGNs have been estimated from
their broad-line widths and optical continuum luminosity (e.g.,
Laor, 2000; McLure \& Dunlop 2001; Gu, Cao, \& Jiang 2001). In
this work, we use the sample of \citet{g01}. The sources in their
sample are selected from a parent sample consisting of 1-Jy, S4,
and S5 radio catalogues. As this is a radio-selected sample, it
includes most strong radio sources. {We assume the physics of BLRs
in radio-loud quasars is similar to radio-quiet quasars in the
sample of \citet{k00}, i.e., the tight correlation between the BLR
size and optical ionizing luminosity is hold for radio-loud
quasars.} The central black hole masses of 86 radio-loud (RL) AGNs
are estimated by using their broad-line width data and optical
continuum luminosities. These 86 AGNs consist of 55 flat-spectrum
sources and 31 steep-spectrum sources. We adopt their sample for
our present investigation. All the profile data of broad emission
line H$\beta$ and derived black hole masses of these sources are
listed in \citet{g01}.

For normal bright AGNs, the bolometric luminosity can be estimated
from their optical luminosity $L_{\lambda,\rm opt}$ at 5100
$\rm\AA$ by \citep{k00} \be L_{\rm bol}\simeq 9\lambda L_{\rm
\lambda,opt}, \label{lbolopt}\ee {which is only a rough estimate.
The coefficient in Eq. (\ref{lbolopt}) may vary for AGNs with
different spectral energy distributions. }

As the central black hole masses are available for the sources in
this sample, we can use Eq. (\ref{lbolopt}) to derive the
dimensionless accretion rates $\dot{m}$ of these AGNs. The derived
dimensionless accretion rate is given by \be
\dot{m}=\dot{M}/\dot{M}_{\rm Edd}\simeq L_{\rm bol}/L_{\rm
Edd}.\label{mdot} \ee This is valid for standard thin accretion
disks, because the radiative efficiency of thin disks is a
constant for given black hole spin parameter $a$. Combining the
relations (\ref{rblr}) and (\ref{mbh}), we can obtain \be M_{\rm
bh}\propto V_{\rm BLR}^2L_{\lambda,\rm
opt}^{0.7}.\label{mbhscale}\ee So, using the relations
(\ref{lbolopt}), (\ref{mdot}), and (\ref{mbhscale}), we get
\be\dot{m}\propto V_{\rm BLR}^{-2} L_{\lambda,\rm
opt}^{0.3},\label{mdot2}\ee since $L_{\rm Edd}\propto M_{\rm bh}$.

{It should be cautious for the estimated black hole masses
especially for flat-spectrum AGNs, because of not having
considered the orientational effects either for the broad-line
width or the beamed optical continuum emission. The fact that the
virilized H$\beta$ black hole mass estimates are systematically
lower than those estimated from $M_{\rm bh}-\sigma$ relation
implies that the BLR geometry may be disk-like. The black hole
masses estimated for flat-spectrum sources using H$\beta$ line
widths may probably be underestimated \citep{md01}. The errors on
the present estimate of the dimensionless accretion rate may be
caused by the uncertainties on the black hole mass estimate
because of orientational  effects, the beamed optical continuum
emission, and the coefficient in relation (\ref{lbolopt}). How
these uncertainties on the estimates of the black hole mass
$M_{\rm bh}$, bolometric luminosity $L_{\rm bol}$, and the
dimensionless accretion rate $\dot{m}$ may affect  our conclusions
will be discussed in Sect. 6. }

\section{Jet power}

The jet power can be estimated from low-frequency radio luminosity
by   \be Q_{\rm jet}\simeq 3\times 10^{38}f^{3/2}L_{\rm ext,151}
^{6/7}~ {\rm W}, \label{qjetrad} \ee where $L_{\rm ext,151}$ is
the extended radio luminosity at 151 MHz in units of 10$^{28}$
W~Hz$^{-1}$~sr$^{-1}$ \citep{w99}. \citet{w99} have argued that
the normalization is very uncertain and introduced the factor $f$
to account for these uncertainties. They use a wide variety of
arguments to suggest that  $1\leq f\leq 20$. In this paper, we
conservatively adopt the lower limit $f=1$. For some flat-spectrum
sources, their radio/optical continuum emission is strongly beamed
towards us because of their relativistic motions and small viewing
angles of the jets with respect to the line of sight. It is found
that the spectra of some flat-spectrum quasars are flat even at
the wavelength around 151 MHz, which implies that the radio
emission from these sources at 151 MHz is still dominated by the
core emission rather than the extended emission. So, the observed
low-frequency radio emission at 151 MHz may still be Doppler
beamed.  We therefore use the extended radio emission measured by
VLA to estimate the jet power (see Cao 2003). The VLA observations
are usually preformed at a frequency much higher than 151 MHz. The
extended radio emission measured by the VLA has to be
$K$-corrected to 151 MHz in the rest frame of the source assuming
$\alpha_{\rm e}=0.8$ ($f_{\nu}\propto \nu^{-\alpha_{\rm e}}$)
\citep{c99}. Apart from steep-spectrum AGNs, we find 41
flat-spectrum AGNs with extended emission data in this sample.

\section{Jet formation mechanisms}

L99 estimated the maximal strength of the large-scale fields
driving the jets, and then the maximal jet power extracted from
the rapidly spinning black hole or the accretion disk on the
assumption of the fields being amplified by dynamo processes.  We
follow L99, the maximal jet power extracted from a rotating black
hole or an accretion disk can be calculated for a standard thin
disk, if the black hole mass $M_{\rm bh}$ and accretion rate
$\dot{m}$ are specified.

\subsection{Jet power extracted from standard thin disks}

The maximal power of the jet accelerated by an magnetized
accretion disk is

\be L_{\rm BP}^{\rm max} =4\pi \int {\frac {B_{\rm pd}^2}{4\pi}}
R^2\Omega(R) dR, \label{lbpsd}\ee where $B_{\rm pd}\sim
B_{\varphi}$ is assumed, and $B_{\rm pd}$ is the strength of the
large-scale ordered poloidal field at the disk surface.

The strength of the field at the disk surface is usually assumed
to scale with the pressure of the disk, as done in \citet{ga97}.
However, L99 pointed out that the large-scale field can be
produced from the small-scale field created by dynamo processes as
$B(\lambda)\propto \lambda^{-1}$ for the idealized case, where
$\lambda$ is the length scale of the field \citep{tp96,r98}. So,
the large-scale field threading the disk is related with the field
produced by dynamo processes can be approximated by (L99)

\be B_{\rm pd}\sim {\frac {H}{R}} B_{\rm dynamo}. \label{mfpd}\ee

The dimensionless pressure scale-height of the disk $H/R$ is given
by \citep{ln89}

\be {\frac {H}{R} } =15.0\dot{m}r^{-1}c_{2}, \label{hr}\ee where
the coefficient $c_{2}$ is defined in Novikov \& Thorne (1973,
hereafter NT73), and the dimensionless quantities are  defined by

\bd r={\frac {R}{R_{\rm G}}},~~~ R_{\rm G}={\frac {GM_{\rm
bh}}{c^2} },~~~ {\dot m}={\frac {\dot M}{\dot M_{\rm Edd}}},~~~
\ed and \be \dot{M}_{\rm Edd}={\frac {L_{\rm Edd}}{\eta_{\rm
eff}c^2}}=1.39\times 10^{15} m~~{\rm kg~s^{-1}},~~~ m={\frac
{M_{\rm bh}}{M_\odot}} \ee where $\eta_{\rm eff}=0.1$ is adopted.

The dimensionless scale-height of the disk $H/R$ is in principle a
function of $R$, and it reaches a maximal value in the inner
region of the disk (Laor \& Netzer 1989). We adopt the maximal
value of $H/R$ in the estimate of large-scale field strength
$B_{\rm pd}$ at the disk surface.

As L99, the strength of the magnetic field produced by dynamo
processes in the disk is given by

\be {\frac {B_{\rm dynamo}^2}{4\pi}} \sim {\frac {W}{2 H}},
\label{mfdyn0} \ee where $W$ is the integrated shear stress of the
disk, and $H$ is the scale-height of the disk. For a relativistic
accretion disk, the integrated shear stress is given by Eq.
(5.6.14a) in NT73. Equation (\ref{mfdyn0}) can be re-written as

\be B_{\rm dynamo}=3.56\times 10^8 r^{-3/4}m^{-1/2} A^{-1}BE^{1/2}
{\rm gauss}, \label{mfdyn} \ee where $A$, $B$ and $E$ are general
relativistic correction factors defined in NT73.

In standard accretion disk models, the angular velocity of the
matter in the disk is usually very close to Keplerian velocity.
For a relativistic accretion disk surrounding a rotating black
hole, the Keplerian angular velocity is given by

\be \Omega(r) =2.034\times 10^5{\frac {1}{m(r^{3/2}+a)} }~{\rm
s}^{-1}, \label{angv} \ee where $a$ is dimensionless specific
angular momentum of a rotating black hole.

We use Eqs. (\ref{mfpd})$-$(\ref{angv}), the maximal power of the
jet accelerated from a magnetized disk is available by integrating
Eq. (\ref{lbpsd}), if some parameters: $m$, $\dot{m}$, $a$, are
specified.

As discussed in L99, the power extracted from a rotating black
hole by the Blandford-Znajek process is determined by the hole
mass $m$, the spin of the hole $a$, and the strength of the
poloidal field threading the horizon of a rotating hole $B_{\rm
ph}$: \be L_{\rm BZ}^{\rm max} = {1 \over 32} \omega_{\rm F}^2
B_\bot^2 R_{\rm h}^2 c
 a^2, \label{lbzsd}
\ee for a black hole of mass $m$ and dimensionless angular
momentum $a$, with a magnetic field $B_\bot$ normal to the horizon
at $R_{\rm h}$. Here the factor $\omega_{\rm F}^2 \equiv
\Omega_{\rm F} (\Omega_{\rm h} - \Omega_{\rm F}) / \Omega_{\rm
h}^2$ depends on the angular velocity of field lines $\Omega_{\rm
F}$ relative to that of the hole, $\Omega_{\rm h}$. In order to
estimate the maximal power extractable from a spinning black hole,
we adopt $\omega_{\rm F} = 1/2$. As the field $B_\bot$ is
maintained by the currents in the accretion disk surrounding the
hole, the strength of $B_\bot$ should be of the same order of that
in the inner edge of the disk, and $B_\bot\simeq B_{\rm pd}(r_{\rm
in})$ is therefore adopted.

\subsection{Jet power extracted from the disk coronas}

In this work, we only consider the case of the jets being
magnetically driven by the fields created in the coronas of the
disks, as recently suggested by \citet{mf02} for the black holes
accreting at low rates.

In the disk-corona scenario, the cold disk and the hot corona
above the disk are in pressure equilibrium. Most gravitational
energy of the accretion matter is released in the hot corona
\citep{hm91,km94,sz94}. A small fraction of the soft photons from
the cold disk is Compton up-scattered to X-ray photons by the hot
electrons in the corona. Even if the magnetic pressure is in
equipartition with the gas pressure, the radiation of the Compton
scattering dominates over the synchrotron radiation in the corona
(e.g., Liu, Mineshige, \& Shibata 2002). Roughly about half of the
scattered X-ray photons illuminate the cold disk
\citep{hm91,no93,c98,k01}. Most soft photons from the disk leave
the system without being scattered in the corona, and they are
observed as optical/UV continuum.  The cold disk in the
disk-corona scenario has roughly about half brightness of a
standard disk without a corona, if they are accreting at the same
rate \citep{hm91}. The gases in the coronas are nearly virilized,
their thermal velocity $V_{\rm th}\sim (GM/R)^{1/2}$. Thus, the
thickness of the corona $H_{\rm c}\sim c_{\rm s}/\Omega_{\rm
K}\sim R$, as the sound speed $V_{\rm s}\sim V_{\rm th}$
\citep{no93}. The maximal magnetic stresses created by the dynamo
processes are \be {\frac {B_{\rm dyn}^2}{4\pi}}\sim {\frac {W_{\rm
c}}{2H_{\rm c}}}, \label{bstress}\ee where $W_{\rm c}=2H_{\rm
c}w_{\rm c}$ is the integrated shear stress of the corona
\citep{ss73,l99}. As the scale-height of the corona $H_{\rm c}\sim
R$, the fields created by the dynamo processes have length scale
of $R$. The maximal jet power can be extracted from the corona in
unit surface area is \be q_{\rm j}^{\rm max}\sim {\frac {B_{\rm
dyn}^2}{4\pi}}R\Omega_{\rm K}(R). \ee The viscous dissipation in
the corona is \be f_{\rm vis}^+= {\frac {1}{2}}W_{\rm
c}R\left|{\frac {d\Omega}{dR}}\right|={\frac {3}{4}}W_{\rm
c}\Omega_{\rm K}(R) \label{fvisplus}\ee in unit surface area
\citep{ss73}. Combing Eqs. (\ref{bstress})-(\ref{fvisplus}) and
noting $H_{\rm c}\sim R$ for the corona, we have \be q_{\rm
j}^{\rm max}\sim {\frac {2}{3}}f_{\rm vis}^+. \label{qjf}\ee
Integrating Eq. (\ref{qjf}) over the the surface of the corona on
the assumption of local equilibrium between the radiation and
energy dissipation in the corona, we obtain \be Q_{\rm jet}^{\rm
max}\sim {\frac {2}{3}} L_{\rm c},\ee where $L_{\rm c}$ is corona
luminosity. In the disk-corona scenario, almost all gravitational
energy of the accretion matter is released in the corona, i.e.,
$L_{\rm c}\simeq L_{\rm bol}$ \citep{hm91}. The emission of the
corona is mainly in X-ray bands, half of which leaves the
disk-corona system and is observed as X-ray emission and the left
half X-ray photons illuminate the cold disk to re-radiate in
optical/UV bands \citep{hm91}. Thus, we have \be L_{\rm X}\simeq
{\frac {L_{\rm bol}}{2}}\ee and \be Q_{\rm jet}^{\rm max}\sim
{\frac {2}{3}}L_{\rm bol}\simeq {\frac {4}{3}}L_{\rm
X},\label{qjetbol} \ee which indicates that the jets can be more
efficiently accelerated from the coronas above the disks than
directly from the thin disks (see the discussion on the jet
accelerated by the fields of the thin disks in Sect. 1).

\section{Spectra of the disks}

The ADAF will transit to a standard optically thick, geometrically
thin accretion disk at a radius outside $R_{\rm tr}$ \citep{e97}.
Here, we consider a general case, i.e., an ADAF is present near
the black hole and it transits to a cold standard disk (SD) beyond
the transition radius $R_{\rm tr}$ \citep{e97}.

The flux due to viscous dissipation in the outer region of the
disk is \be F_{\rm vis}(R)\simeq {\frac {3GM_{\rm bh}\dot M}{8\pi
R^3}}, \label{fvis} \ee which is a good approximation for $R_{\rm
tr}\gg R_{\rm in}$. The local disk temperature of the thin cold
disk is \be T_{\rm disk}(R)={\frac {F_{\rm
vis}^{1/4}(R)}{\sigma_{\rm B}^{1/4}}}, \label{tdisk} \ee by
assuming local blackbody emission. In order to calculate the disk
spectrum, we include an empirical color correction for the disk
thermal emission as a function of radius. The correction has the
form \citep{chiang02} \be f_{\rm col}(T_{\rm disk}) = f_\infty -
\frac{(f_\infty - 1) [1 +
                     \exp(-\nu_{\rm b}/\Delta\nu)]} { 1 +
                     \exp[(\nu_{\rm p} -\nu_{\rm b})/\Delta\nu]}, \label{fcol}
\ee where $\nu_p \equiv 2.82k_B T_{\rm disk}/h$ is the peak
frequency of blackbody emission with temperature $T_{\rm disk}$.
This expression for $f_{\rm col}$ goes from unity at low
temperatures to $f_\infty$ at high temperatures with a transition
at $\nu_{\rm b} \approx \nu_{\rm p}$. \citet{chiang02} found  that
$f_\infty = 2.3$ and $\nu_b = \Delta\nu = 5\times 10^{15}$\,Hz can
well reproduce the model disk spectra of \citet{h01}. The disk
spectra can therefore be calculated by \be L_\nu =8\pi^2 \left(
{\frac {GM}{c^2}} \right)^2 {\frac{h \nu^3}{c^2 } }
     \int\limits_{r_{\rm tr}}^\infty
     {\frac{r dr}{f_{\rm col}^4[\exp(h\nu/f_{\rm col} k_B T_{\rm disk}) - 1]}}.
\ee In this ADAF$+$SD scenario, the ionizing luminosity from the
disk is a combination of the emission from the inner ADAF and
outer standard disk regions. \citet{cao02a}'s calculations
indicate that the optical continuum emission from the inner ADAF
region can be neglected compared with that from outer SD region if
$R_{\rm tr}$ is around tens to several hundreds Schwarschild
radii, which is due to low radiation efficiency of ADAFs.

\section{Results}

In Fig. \ref{fig1}, we plot the relation between $L_{\rm
bol}/L_{\rm Edd}$ and $Q_{\rm jet}/L_{\rm bol}$. The jet power of
AGNs is estimated from the low-frequency extended radio luminosity
by using relation (\ref{qjetrad}). The maximal jet power can be
extracted by the magnetic fields created in  the standard
accretion disks (solid line) and the rapidly spinning black holes
of $a=0.95$ (dotted line) are calculated by using the method
described in Sect. 4.1. The flat-spectrum and steep-spectrum
sources are labelled as squares and circles respectively. The
sources above the solid line in Fig. \ref{fig1} are labelled as
filled square(circles) (hereafter these sources are referred as
high-jet-power sources). This indicates that the Blandford-Payne
mechanism is unable to produce sufficient jet power observed in
the high-jet-power sources, if only standard thin accretion disks
are present in these sources. The jets in almost all sources of
this sample cannot be powered only by the Blandford-Znajek
mechanism, even if the black holes in these sources are rapidly
spinning at $a=0.95$.

\figurenum{1}
\centerline{\includegraphics[angle=0,width=10.0cm]{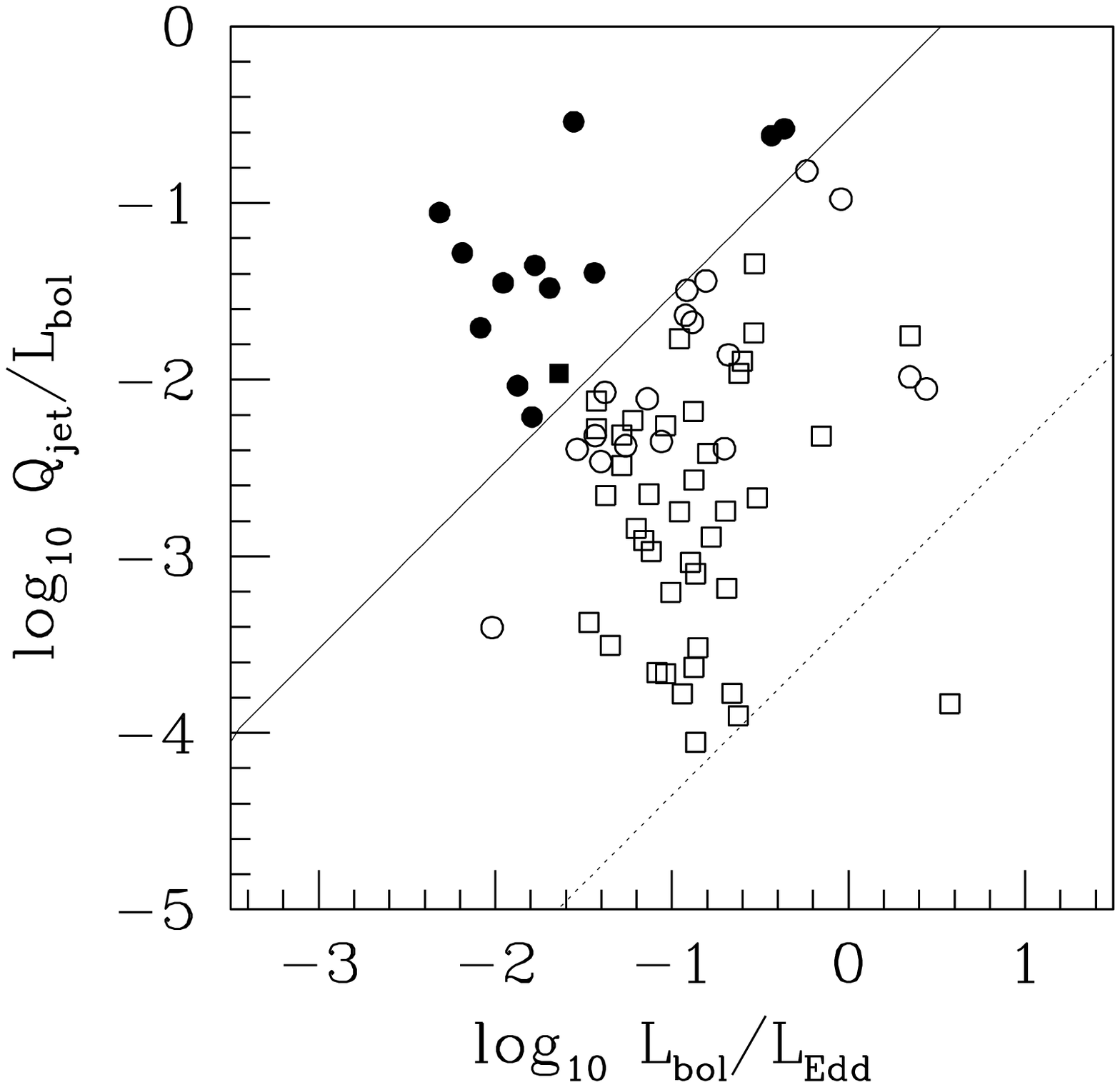}}
\figcaption{\footnotesize The ratio $L_{\rm bol}/L_{\rm Edd}$
versus the ratio $Q_{\rm jet}/L_{\rm bol}$. The squares represent
flat-spectrum sources, and the circles are for steep-spectrum
sources.  The solid line represents the maximal jet power
extracted from a standard thin accretion disk (the Blandford-Payne
mechanism), while the dotted line represents the maximal jet power
extracted from a rapidly spinning black hole $a=0.95$ surrounded
by a standard thin accretion disk (the Blandford-Znajek
mechanism). The sources above the solid line, referred as
high-jet-power sources, are labelled as filled circles(square).
\label{fig1}} \centerline{}

{As mentioned in Sect. 2, there are uncertainties on the estimates
of the black hole mass $M_{\rm bh}$, bolometric luminosity $L_{\rm
bol}$, and then the dimensionless accretion rate $\dot{m}$. The
beamed optical continuum emission from the jets may lead to
overestimate of bolometric luminosity $L_{\rm bol}$ especially for
flat-spectrum sources, which would shift the real locations of the
sources in Fig. \ref{fig1} towards the direction away from the
line (solid) representing the maximal jet power extracted by the
Blandford-Payne mechanism (see Fig. \ref{fig1}). If the disk-like
BLR geometry is present in these sources, their black hole masses
may be underestimated \citep{md01} and the real locations of the
sources should also be shifted towards the direction away from the
maximal jet power line in Fig. \ref{fig1}. The uncertainty on the
coefficient in relation (\ref{lbolopt}) can also lead to errors on
the estimate of bolometric luminosity $L_{\rm bol}$. If the
present coefficient in relation (\ref{lbolopt}) is overestimated,
the locations of high-jet-power sources will be also shifted
towards the direction away from the maximal jet power line. The
jets in all these high-jet-power sources can be extracted by the
Blandford-Payne mechanisms for thin disk cases, only if the
coefficient in relation (\ref{lbolopt}) is significantly
underestimated at more than one order of magnitude, i.e., $>100$
is required instead of 9 in relation (\ref{lbolopt}).  So, the
conclusion that the jets in these high-jet-power sources cannot be
accelerated by the Blandford-Payne mechanism for standard thin
disk cases will not be altered by the uncertainties of the
estimates on the black hole mass $M_{\rm bh}$, bolometric
luminosity $L_{\rm bol}$, or dimensionless accretion rate
$\dot{m}$, unless  the present coefficient in relation
(\ref{lbolopt}) is underestimated at an order of magnitude, which
seems impossible. }

The relation between black hole mass $M_{\rm bh}$ and optical
continuum luminosity $L_\lambda$ at 5100 $\AA$ is plotted in Fig.
\ref{fig3}. The optical continuum emission from a pure ADAF can be
calculated by using the approach proposed by \citet{m97}. We use
the same approach proposed by \citet{cao02a}  to calculate the
maximal optical continuum emission from an ADAF as a function of
black hole mass $M_{\rm bh}$. The maximal optical continuum
emission requires the parameter $\beta=0.5$, which describes the
magnetic field strength with respect to gas pressure, and
viscosity $\alpha=1$. Changing the value of accretion rate
$\dot{m}$, we can find the maximal optical continuum luminosity
for given black hole mass (see Cao, 2002a for details). It is
found that all sources have optical continuum luminosity higher
than the maximal optical luminosity expected from pure ADAFs,
which implies that the emission from pure-ADAFs is unable to
explain the optical ionizing luminosity of these sources. We then
consider another possibility, i.e., the ADAF transits to a
standard thin disk outside the transition radius $R_{\rm tr}$. The
theoretical calculations of the optical continuum luminosity for
such ADAF$+$SD systems with different accretion rates and
transition radii are also plotted in the figure (the calculations
are carried out as described in Sect. 5). It is found that ADAFs
may be present in the inner region of the disk in these
high-jet-power sources only if the critical accretion rates
$\dot{m}_{\rm crit}$ are as high as $\ge$0.05. {For flat-spectrum
sources, the optical continuum emission from the accretion disks
may be overestimated and the accretion rates inferred from Fig.
\ref{fig3} may be larger than their real values. The beaming
effect is believed to be  unimportant for steep-spectrum sources
because of their relatively larger angles than flat-spectrum
sources with respect to the line of sight, and it is therefore
suggested that the optical continuum emission of steep-spectrum
sources is mainly from their accretion disks \citep{s98}. The
high-jet-power sources in this sample are steep-spectrum sources
except one flat-spectrum source. So, the estimated dimensionless
accretion rates for these high-jet-power sources should be less
affected by the beamed jet emission than flat-spectrum sources,
and the main results on the accretion rates and transition radii
for these high-jet-power sources should not be changed
significantly. }

\figurenum{2}
\centerline{\includegraphics[angle=0,width=10.0cm]{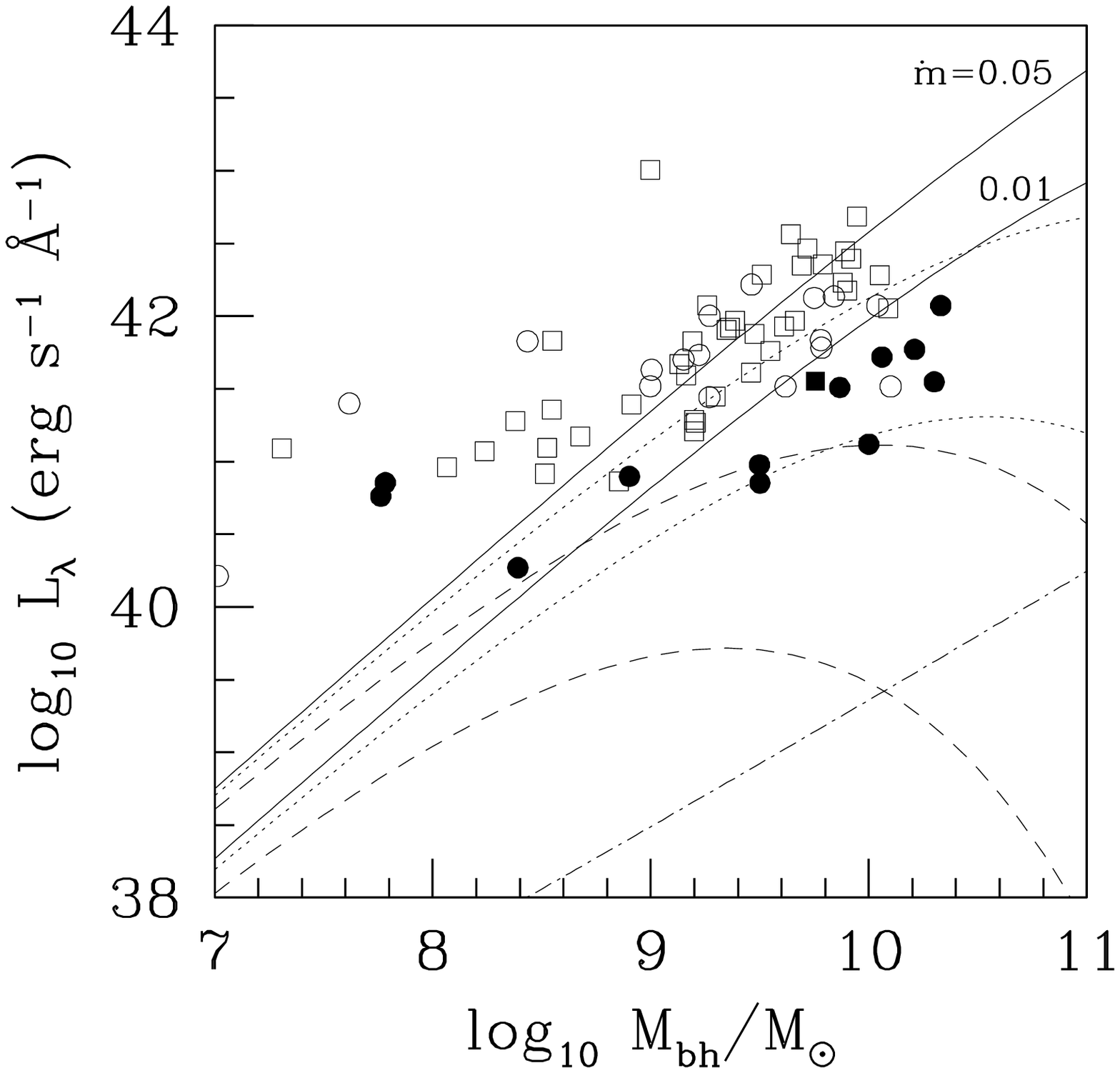}}
\figcaption{\footnotesize The black hole mass $M_{\rm bh}$ versus
optical luminosity $L_{\lambda}$ at 5100~\AA. The dash-dotted line
represents the maximal optical continuum luminosity from
pure-ADAFs. The solid lines represent standard disks accreting at
$\dot{m}=0.01$ and 0.05, respectively. The dotted and dashed lines
represent the ADAF$+$SD models with different transition radii
$R_{\rm tr}=20GM/c^2$ and $50GM/c^2$, respectively. The upper
lines are plotted for $\dot{m}=0.05$, while the lower lines are
for $\dot{m}=0.01$. The symbols are same as Fig. \ref{fig1}.
\label{fig3}} \centerline{}

The relation between bolometric luminosity $L_{\rm bol}$ and jet
power $Q_{\rm jet}$ is plotted in Fig. \ref{fig4}.  The jet power
is lower than the bolometric luminosity for  all sources in this
sample, which would still hold even if the bolometric luminosity
is overestimated by less than an order of magnitude for
flat-spectrum sources.

In order to test relation (\ref{qjetbol}) between $Q_{\rm
jet}^{\rm max}$ and total X-ray luminosity $L_{\rm X}$ from the
disk coronas predicted by the scenario of the jets being
magnetically accelerated by the fields created in the disk
coronas, we roughly convert X-ray luminosity in $0.1-2.4$~keV
$L_{\rm X,0.1-2.4keV}$ detected by ROSAT \citep{v99} to the total
X-ray luminosity assuming the X-ray continuum emission to be in
the range from $0.01-100$~keV with a mean energy spectral index
$\alpha_{\rm X}=1$ ($f_{\rm X}\propto E^{-\alpha_{\rm X}}$). This
is consistent with the general features of  the theoretical
calculations on the X-ray spectra from the disk coronas (e.g.,
Nakamura \& Osaki 1993). In this case, the total X-ray luminosity
$L_{\rm X}\simeq 2.9 L_{\rm X,0.1-2.4keV}$. In Fig. \ref{fig5},
the relation between $K$-corrected core radio luminosity $L_{\rm
c,5G}$ and X-ray luminosity $L_{\rm X}$. The linear regression
considering errors in both coordinates gives \citep{p92} \be
\log_{10} L_{\rm X} =0.866\log_{10}L_{\rm c,5G}+7.517 \ee for 40
flat-spectrum sources (the solid line in Fig. \ref{fig5}), and \be
\log_{10} L_{\rm X} =0.371\log_{10}L_{\rm c,5G}+29.272  \ee for 18
steep-spectrum sources (the dotted line in Fig. \ref{fig5}). The
Spearmann correlation analyses \citep{p92} show that the
correlation for flat-spectrum sources is at the significant level
of 99.98 per cent, while it  becomes 98.5 per cent for
steep-spectrum sources. The different slopes of the correlations
between $L_{\rm X}$ and $L_{\rm c,5G}$ for steep-spectrum and
flat-spectrum sources are obvious, which  may imply that the
observed X-ray emission from these two kinds of AGNs may have
different origins.

\figurenum{3}
\centerline{\includegraphics[angle=0,width=10.0cm]{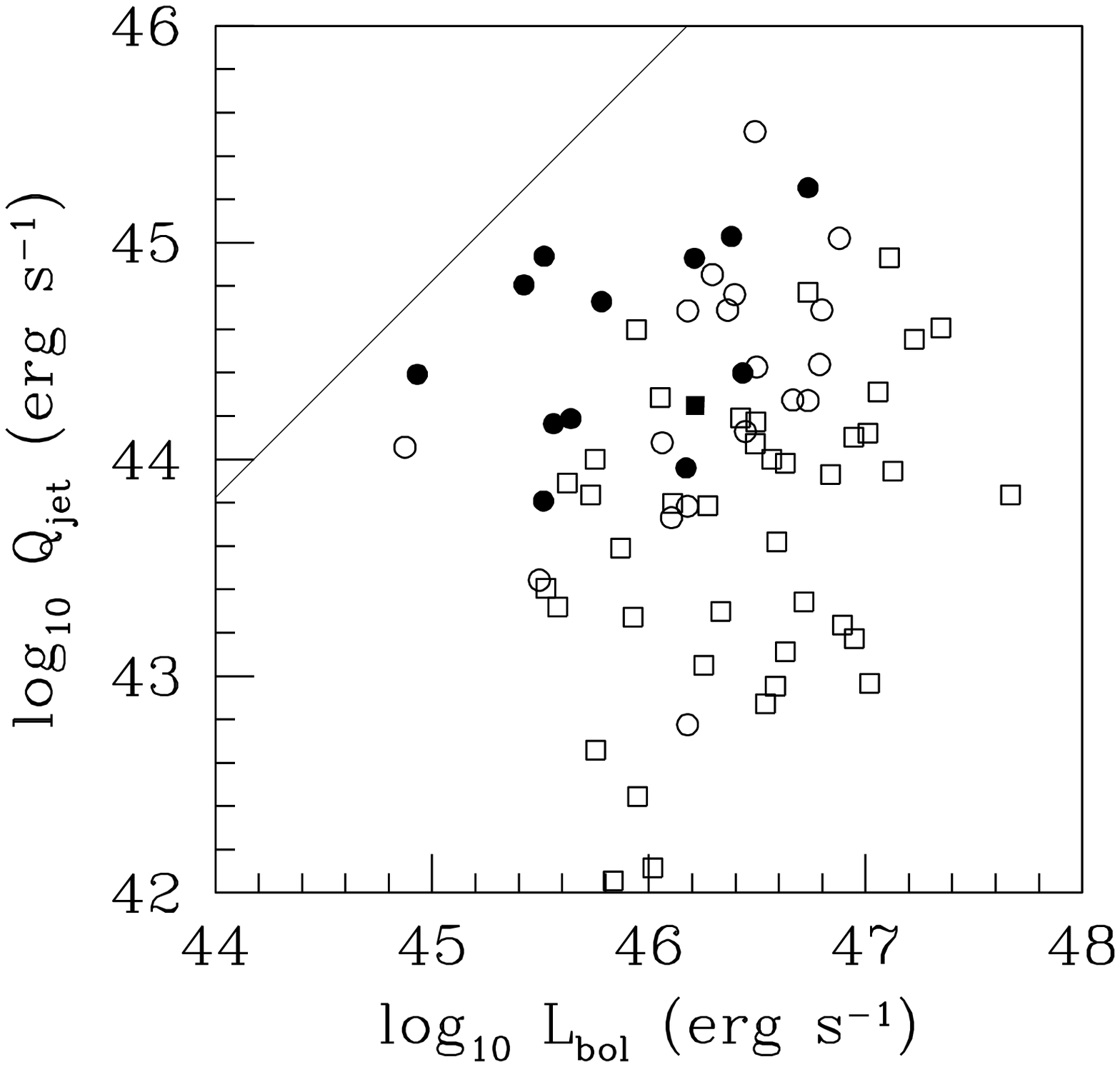}}
\figcaption{\footnotesize
 The relation between bolometric luminosity $L_{\rm
bol}$ and jet power $Q_{\rm jet}$. The line represents $Q_{\rm
jet}=2/3~L_{\rm bol}$. The symbols are same as Fig. \ref{fig1}.
\label{fig4}}\centerline{}

\figurenum{4}
\centerline{\includegraphics[angle=0,width=10.0cm]{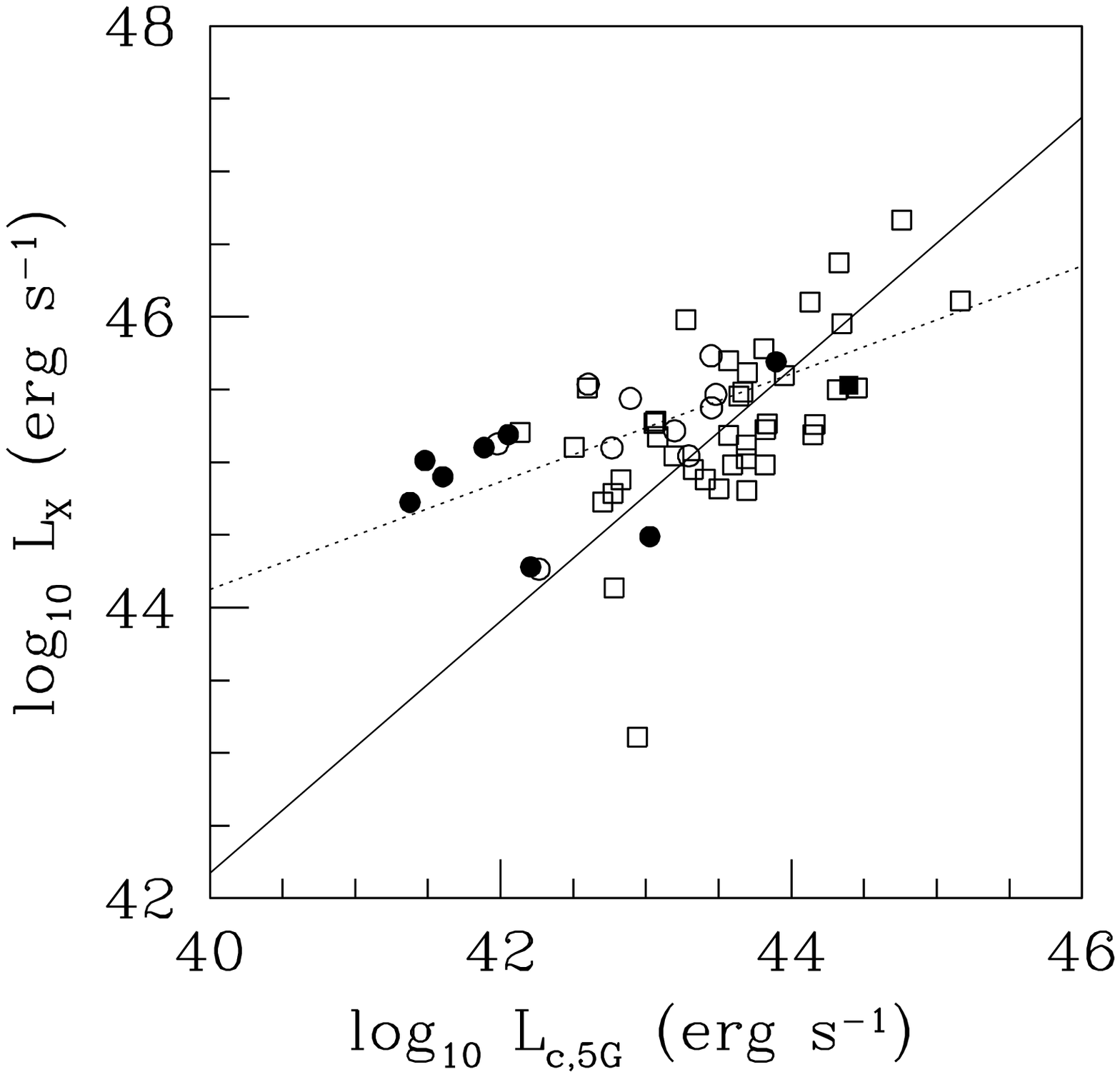}}
\figcaption{\footnotesize The relation between X-ray luminosity
$L_{\rm X}$ and core radio luminosity $L_{\rm c}$ at 5 GHz. The
solid and dotted lines represent the linear regressions for
flat-spectrum and steep-spectrum sources respectively. The symbols
are same as Fig. \ref{fig1}. \label{fig5}}\centerline{}

We plot the relation between X-ray luminosity $L_{\rm X}$ and jet
power $Q_{\rm jet}$ in Fig. \ref{fig6}. We find that almost all
sources have $Q_{\rm jet}<L_{\rm X}$, except one steep-spectrum
source with $Q_{\rm jet}\sim L_{\rm X}$.

\figurenum{5}
\centerline{\includegraphics[angle=0,width=10.0cm]{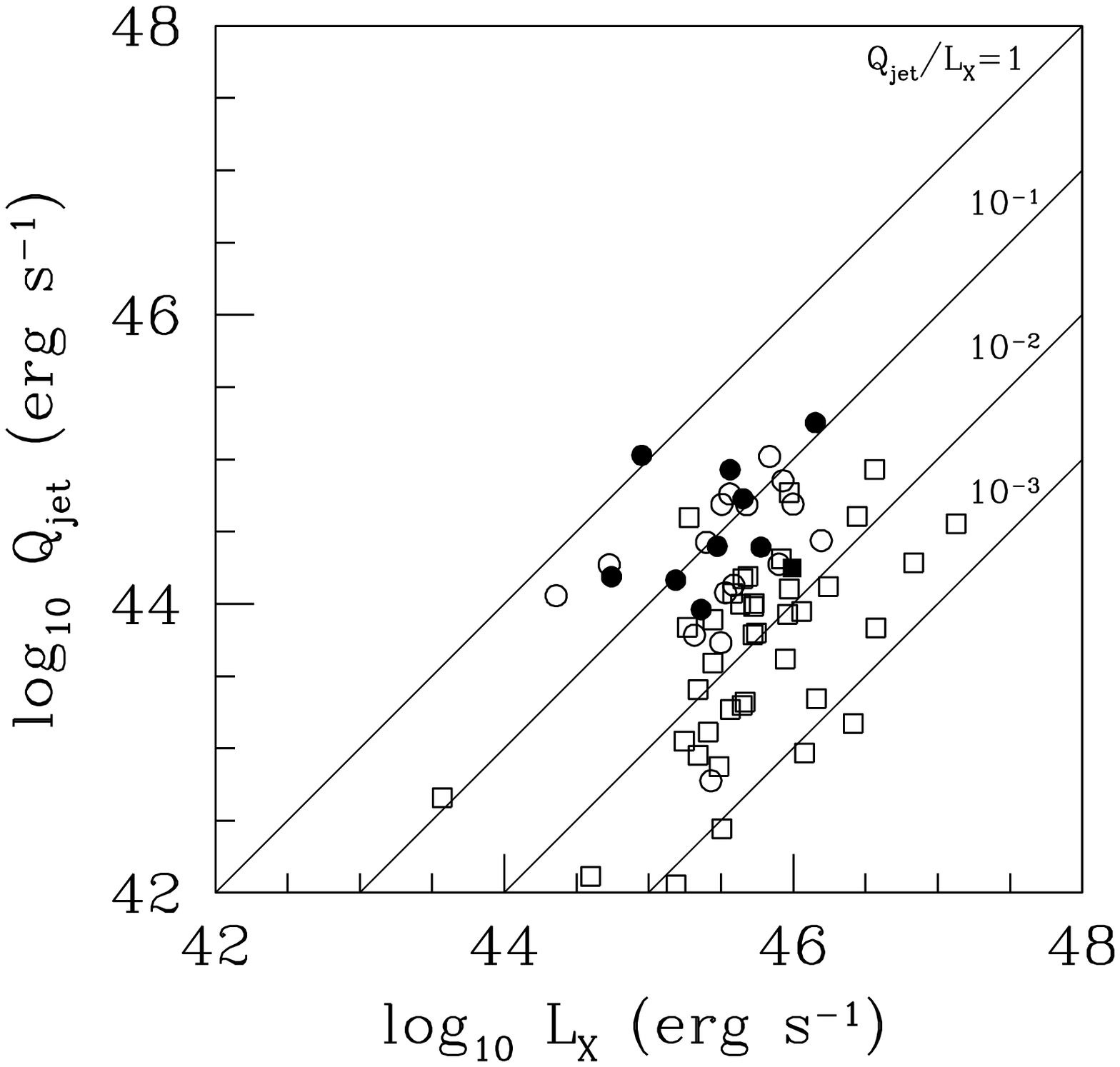}}
\figcaption{\footnotesize The relation between X-ray luminosity
$L_{\rm X}$ and jet power $Q_{\rm jet}$. The lines represent
$Q_{\rm jet}/L_{\rm X}=10^{-3}$, $10^{-2}$, $10^{-1}$, and 1,
respectively. The symbols are same as Fig.
\ref{fig1}.\label{fig6}}\centerline{}

\section{Discussion}

There are 13 sources with jet power above the maximal jet power
expected to be extracted from standard thin disks (see Fig.
\ref{fig1}), i.e., the fields of thin disks are unable to drive
such strong jets in these high-jet-power sources. The jet can be
accelerated from an ADAF more efficiently than a standard thin
disk, though an accurate calculation on the maximal jet power
extracted from an ADAF is still unavailable  \citep{m01}.
\citet{m01} pointed out that general relativistic effects may be
important and a rough estimate of jet power shows that the maximal
jet power extracted from an ADAF as high as $10^{45-46}$
erg~s$^{-1}$ may be possible for a rapidly spinning massive black
hole ($\sim 10^9 M_\odot$). If ADAFs are present in these sources,
they will have fainter optical continuum emission than standard
disks due to the lower accretion rates and lower radiation
efficiency of ADAFs. It is found that pure-ADAFs are unable to
produce observed bright optical continuum luminosity for all
sources in this sample (see Fig. \ref{fig3}). However, this cannot
rule out the presence of ADAFs in these sources, because the ADAF
may transit to a standard thin disk in the region outside the
transition radius $R_{\rm tr}$. In this case, the standard thin
disk in the outer region may be responsible for the observed
optical continuum emission. The ADAF$+$SD (standard disk) scenario
is tested against observational optical continuum luminosity in
Fig. \ref{fig3}. As 12 of 13 high-jet-power sources are
steep-spectrum AGNs, the optical continuum emission is believed to
be mainly from the accretion disks \citep{s98}. It is found that
relatively high accretion rates $\dot{m}\ge 0.05$ are required to
explain the optical continuum luminosity of these high-jet-power
sources, if the ADAFs are truncated at rather small radii (e.g.,
$\le 20$ Schwarschild radii). If the transition radii are larger
than this value, even higher accretion rates are required. The
exact value of the critical accretion rate $\dot m_{\rm crit}$ is
still unclear for an ADAF, which depends on the value of the disk
viscosity parameter $\alpha$, i.e., $\dot{m}_{\rm crit}\simeq
0.28\alpha^2$ \citep{m97}. Recent three-dimensional MHD
simulations suggest that the viscosity $\alpha$ in the discs to be
$\sim 0.1$ \citep{a98}, or $\sim 0.05-0.2$ \citep{hb02}.  If a
value $\alpha= 0.2$ is conservatively adopted, the accretion rates
$\dot{m}\le \dot m_{\rm crit}\sim 0.01$ are required for ADAFs.
This may probably imply that ADAFs are not present in these
high-jet-power sources.

For an ADAF, its bolometric luminosity can be significantly lower
than the jet power extracted from the ADAF or the rapidly spinning
black hole surrounded by the ADAF, because the radiation
efficiency of ADAFs is much lower than the efficiency of power
channelled into the jet \citep{an99,m01}. We have not found any
source in our sample with $Q_{\rm jet}>L_{\rm bol}$ (see Fig.
\ref{fig4}). It  implies that standard thin disks, at least in the
outer regions of the disks, are required to produce observed
bright bolometric luminosity, even if ADAFs are present in these
sources in the case of $\alpha >0.2$, though it seems inconsistent
with MHD simulations \citep{hb02}.

{The ADAF solutions may be modified to adiabatic inflow-outflow
solutions (ADIOSs), if a powerful  wind is present to carry away
mass, angular momentum and energy from the accreting gas
\citep{bb99}. In this case, the accretion rate of the disk is a
function of radius $r$ instead of a constant accretion rate along
$r$ for a pure ADAF. For an ADIOS flow, the gas swallowed by the
black hole is only a small fraction of the rate at which it is
supplied, as most of the gas is carried away in the wind before it
reaches the black hole. The ADIOS flow has similar local structure
as the  ADAF accreting at the same rate, so the accretion rate
$\dot{m}$ for an ADIOS flow at any radius should be smaller than
the critical rate $\dot{m}_{\rm crit}$ of ADAFs. The ADIOS flow is
fainter than an ADAF, if they are accreting at the same rate at
the outer radius, because the accretion rate of an ADIOS flow
decreases while the gas is flowing onto the hole (e.g.,
\citet{qn99}; \citet{ccy02}). So, even if an ADIOS flow is present
instead of an ADAF in the inner region of the disk, the optical
continuum emission is still dominated by the emission from the
standard thin disk in the outer region, as the ADAF case
\citep{cao02a}. Our present spectral analysis for ADAF$+$SD system
are valid even for ADIOS$+$SD systems.  For convection dominated
accretion flows (CDAFs), most of the gas circulates in convection
eddies rather than accreting onto the black hole \citep{n00,qg00}.
The accretion rates of CDAFs are much smaller than non-convecting
ADAFs, and CDAFs are very faint \citep{b01}. This implies that the
accretion rate in a standard thin disk surrounding a CDAF should
be much smaller than the critical accretion rate $\dot{m}_{\rm
crit}$ in ADAFs, if a steady accretion flow is assumed. The
observed bright optical continuum emission in these high-jet-power
sources requires the black holes accreting at rather high rates,
which seems to rule out the possibility of CDAFs surrounded by
standard thin disks in these high-jet-power  sources.  }

{The powerful  jets in these high-jet-power sources may be
accelerated from ADAFs (or ADIOS flows) surrounded by standard
thin disks outside $\sim$ 20 Schwarzschild radii accreting at
$\dot{m}\ge 0.05$, which is required by observed bright optical
continuum emission in these sources. However, such high accretion
rates require a high viscosity $\alpha$, which seems inconsistent
with MHD simulations \citep{hb02}. } Here, we propose an
alternative explanation, i.e., the jets, at least in these
high-jet-power sources, are accelerated from the coronas of the
disks. In this case, the maximal power of the jet accelerated from
the corona of the disk can be as high as its bolometric luminosity
(see Eq. (\ref{qjetbol}) and Fig. \ref{fig4}), and the strong jets
in these high-jet-power sources can be naturally explained by this
scenario. The cold disk irradiated by the corona above can produce
bright optical continuum emission as observed in these
high-jet-power sources.

It should be cautious that the calculations carried out in Sect.
4.2 are on the assumption of a perfect parallel corona structure.
The real corona above the cold disk may not always have perfect
parallel structure. The time variations and spatial inhomogeneity
can be caused by the magnetic fields (e.g., Kawaguchi et al.
2000). So, the maximal jet power extractable from the corona may
be lower than that estimated by Eq. (\ref{qjetbol}) because of the
inhomogeneous corona structure. Most high-jet-power sources in our
sample have $Q_{\rm jet}/L_{\rm bol}\le 0.1$, while three sources
have $Q{\rm jet}/L_{\rm bol}\simeq 0.3$ (see Fig. \ref{fig1}). The
disk-corona scenario may still be a reasonable explanation for jet
formation in these high-jet-power sources.

{The emission from the coronas is mainly in X-ray bands, and this
scenario predicts a relation between the maximal power of the jet
accelerated from the corona and X-ray luminosity of the corona
(Eq. \ref{qjetbol}). This prediction can be tested against
observations. However, the difficulty arises from the fact that
the observed X-ray emission from these radio-loud AGNs may be a
mixture of the emission from the jets and accretion disk coronas.
If the X-ray emission from the jets in radio-loud AGNs dominates
over that from the disk coronas, a correlation between radio core
emission and X-ray emission is expected, because both of them are
from the jets and Doppler beamed. Figure \ref{fig5} gives
different correlations between core radio luminosity $L_{\rm
c,5G}$ and X-ray luminosity $L_{\rm X}$ for flat-spectrum and
steep-spectrum AGNs. \citet{fa03} used LINERs, FR I galaxies, and
BL Lac objects in their investigations on the radio-X-ray
correlation, in which ADAFs are believed to be present and  the
radio and X-ray emission are believed to be dominated by the
emission from the jets. The correlation found in this work between
radio and X-ray luminosities for flat-spectrum sources is roughly
consistent with the correlation given by \citet{fa03}. The core
dominance parameter $R_{\rm c}$ varies over several orders of
magnitude for the sources in this sample \citep{cj01}, which
reflects the Doppler factors spread over a large range for these
sources. In this work, we do not intend to explore this
correlation in detail as done by \citet{fa03}, because of unknown
Doppler beaming factors of  the sources in this sample.
Nevertheless, the different slopes of the correlations for
flat-spectrum and steep-spectrum sources do indicate the different
origins of the X-ray emission to be in these two kinds of sources.
Compared with flat-spectrum sources, most steep-spectrum sources
have brighter X-ray emission than their flat-spectrum counterparts
with similar radio core emission (see Fig. \ref{fig5}). This may
imply that the X-ray emission from the disk coronas rather than
the jets is dominant for steep-spectrum AGNs. If this is the case,
the observed X-ray emission from steep-spectrum sources is
probably from the disk coronas. We can tentatively use Fig.
\ref{fig6} to test our estimate on the relation between $Q_{\rm
jet}^{\rm max}$ and $L_{\rm X}$ (Eq. (\ref{qjetbol})) expected for
the jets magnetically accelerated from the coronas of the disks.
It is found in Fig. \ref{fig6} that $Q_{\rm jet}<L_{\rm X}$ is
satisfied for almost all sources except one with $Q_{\rm jet}\sim
L_{\rm X}$, which is consistent with the scenario of the jets
being magnetically accelerated from the disk coronas. If this
scenario is correct, it implies that the factor $f$ in Eq.
(\ref{qjetrad}) describing the uncertainty of the jet power
estimated from the extended radio luminosity should be close to
unit at least for these high-jet-power sources (see Figs.
\ref{fig4} and \ref{fig6}). }

Recently, \citet{huj02} proposed that the jet is launched from a
layer, governed by a highly diffusive, super-Keplerian rotating
and thermally dominated by viril-hot and magnetized ion-plasma.
This layer is located between the accretion disk and the corona
surrounding the nucleus. In this model, most of the accretion
energy can be converted into magnetic and kinetic energies that go
into powering the jet \citep{huj03,huj04}, and the powerful jet
with $Q_{\rm jet}\sim L_{\rm bol}$ can form naturally in such a
layer. This model, in principle, can explain the powerful jets of
the high-jet-power sources in our present sample, though the
detailed numerical model calculations for the jets in these
sources are beyond the scope of this paper.

\acknowledgments This work is supported by the National Science
Fund for Distinguished Young Scholars (grant 10325314), NSFC
(grants 10173016; 10333020), and the NKBRSF (grant G1999075403).
This research has made use of the NASA/IPAC Extragalactic Database
(NED), which is operated by the Jet Propulsion Laboratory,
California Institute of Technology, under contract with the
National Aeronautic and Space Administration.


\begin{thebibliography}{}

\bibitem[Armitage(1998)]{a98}
  Armitage P.J., 1998, ApJ, 501, L189
\bibitem[Armitage \& Natarajan(1999)]{an99}
  Armitage P.J., \& Natarajan P., 1999, \apj, 523, L7
\bibitem[Ball, Narayan \& Quataert(2001)]{b01}
  Ball, G. H., Narayan, R., Quataert, E., 2001, \apj, 552, 221
\bibitem[Blandford \& Begelman(1999)]{bb99}
  Blandford R.D., \& Begelman M.C., 1999, \mnras, 303, L1
\bibitem[Blandford \& Payne(1982)]{bp82}
   Blandford R. D., \& Payne D. G., 1982, \mnras, 199, 883
\bibitem[Blandford \& Znajek(1977)]{bz77}
   Blandford, R. D., \& Znajek, R. L., 1977, \mnras, 179, 433
\bibitem[Cao(2002a)]{cao02a}
       Cao, X., 2002a, \apj, 570, L13
 \bibitem[Cao(2002b)]{cao02b}
      Cao, X., 2002b, \mnras, 332, 999
\bibitem[Cao(2003)]{cao03}
      Cao, X., 2003, \apj, 599, 147
\bibitem[Cao \& Jiang(1999)]{cj99}
      Cao, X., Jiang, D.R., 1999, \mnras, 307, 802
\bibitem[Cao \& Jiang(2001)]{cj01}
      Cao, X., Jiang, D.R., 2001, \mnras, 320, 347
\bibitem[Cao et al.(1998)]{c98}
      Cao, X., Jiang, D.R., You, J.H., Zhao, J.L., 1998, \aap,
      330, 464
\bibitem[Cassaro et al.(1999)]{c99}
   Cassaro, P., Stanghellini, C., Bondi, M., Dallacasa, D., della Ceca, R., \& Zappal¨¤, R.
   A., \aaps, 139, 601
\bibitem[Chang, Choi \& Yi(2002)]{ccy02}
  Chang H.Y., Choi C.S., \& Yi I., 2002, \aj, 124, 1948
\bibitem[Chiang(2002)]{chiang02}
   Chiang, J., 2002, \apj, 572, 79
\bibitem[Esin, McClintock, \&  Narayan(1997)]{e97}
     Esin, A. A., McClintock, J. E., \&  Narayan, R., 1997, \apj, 489, 865
\bibitem[Falcke \& Biermann(1995)]{fb95}
     Falcke, H., \& Biermann, P.L., 1995, \aap, 293, 665
\bibitem[Falcke, Koerding, \& Markoff(2004)]{fa03}
     Falcke, H., K\"ording, E., \& Markoff, S.,  2004, \aap, 414, 895
 \bibitem[Ferrarese \& Merritt(2000)]{fm00}
    Ferrarese, L.,  \& Merritt, D., 2000, \apj, 539, L9
\bibitem[Gebhardt et al.(2000)]{g00}
    Gebhardt, K. et al., 2000, \apj, 539, L13
\bibitem[Ghosh \& Abramowicz(1997)]{ga97}
     Ghosh P., \& Abramowicz M. A., 1997, \mnras, 292, 887
\bibitem[Gu, Cao, \& Jiang(2001)]{g01}
     Gu, M., Cao, X., \& Jiang, D.R., 2001, \mnras, 327, 1111
\bibitem[Haardt \& Maraschi(1991)]{hm91}
     Haardt, F., \& Maraschi, L., 1991, \apj, 380, L51
\bibitem[\protect\citeauthoryear{Hawley \& Balbus}{2002}]{hb02}
  Hawley, J.F., Balbus, S.A., 2002, ApJ, 573, 738
\bibitem[Hubeny et al.(2001)]{h01}
     Hubeny, I., Blaes, O., \& Krolik, J. H.,  Agol, E., \apj, 559,
     680
\bibitem[Hujeirat(2004)]{huj04}
     Hujeirat, A., 2004, \aap, 416, 423
\bibitem[Hujeirat, Camenzind, \& Livio(2002)]{huj02}
     Hujeirat, A., Camenzind, M., \& Livio, M., 2002, \aap, 394, L9
\bibitem[Hujeirat et al.(2003)]{huj03}
     Hujeirat, A., Livio, M., Camenzind, M., \& Burkert, A., 2003,
     \aap, 408, 415
\bibitem[Kaspi et al.(2000)]{k00}
  Kaspi S., Smith P.S., Netzer H., Maoz D., Jannuzi B.T., \& Giveon
  U., 2000, ApJ, 533, 631
\bibitem[Kawaguchi et al.(2000)]{ka00}
     Kawaguchi, T., Mineshige, S., Machida, M., Matsumoto,
R., Shibata, K., 2000, \pasj, 52, L1
\bibitem[Kawaguchi, Shimura, \& Mineshige(2001)]{k01}
  Kawaguchi, T., Shimura, T., Mineshige, S., 2001, \apj, 546. 966
\bibitem[Kusunose, \& Mineshige(1994)]{km94}
    Kusunose, M., \& Mineshige, S., 1994, \apj, 423, 600
\bibitem[Laor(2000)]{l00}
    Laor, A., 2000, \apj, 543, L111
\bibitem[Laor \& Netzer(1989)]{ln89}
     Laor A., \& Netzer H., 1989, \mnras, 238, 897
 \bibitem[Liu, Mineshige, \& Shibata(2002)]{liu02}
     Liu, B.F., Mineshige, S., \& Shibata, K., 2002, \apj, 572,
     L173
\bibitem[Livio, Ogilivie, \& Pringle(1999)]{l99}
     Livio M., Ogilivie G. I., \& Pringle J. E., 1999, \apj, 512, 100(L99)
\bibitem[Mahadevan(1997)]{m97}Mahadevan, R., 1997, \apj, 477, 585
\bibitem[McLure \& Dunlop(2001)]{md01}
     McLure, R. J., \& Dunlop, J. S., 2001, \mnras, 327, 199
\bibitem[McLure \& Dunlop(2002)]{md02}
        McLure, R.J., \& Dunlop, J.S., 2002, \mnras. 331, 795
\bibitem[Meier(2001)]{m01}
     Meier, D. L., 2001, \apj, 548, L9
\bibitem[Merloni \& Fabian(2002)]{mf02}
     Merloni, A., Fabian, A. C., 2002, \mnras, 332, 165
\bibitem[Merloni, Heinz, \& Di Matteo(2003)]{m03}
     Merloni, A., Heinz, S., \& Di Matteo, T., 2003, \mnras, 345,
     1057
\bibitem[Nakamura \& Osaki(1993)]{no93}
    Nakamura, K.,  \& Osaki, Y., 1993, \pasj, 45, 775
\bibitem[Narayan, Igumenshchev, \& Abramowicz(2000)]{n00}
   Narayan, R., Igumenshchev, I.V., \& Abramowicz, M.A., 2000, \apj, 539, 798
\bibitem[Novikov \& Throne (1973)]{nt73}
    Novikov I., \& Throne K. S., 1973, in Black holes, eds de Witt C. and
    de Witt B., Gordon \& Breach, New York(NT73)
\bibitem[Peterson(1993)]{p93}
    Peterson, B. M., 1993, \pasp, 105, 247
\bibitem[Press et al.(1992)]{p92}
    Press, W.H., Teukolsky, S.A., Vetterling W.T., \& Flannery B.P., 1992,
    Numerical recipes in FORTRAN: the art of Scientific computing, Cambridge
    university press.
\bibitem[Quataert \& Gruzinov(2000)]{qg00}
   Quataert, E., \&  Gruzinov, A., 2000, \apj, 539, 809
\bibitem[Quataert \& Narayan(1999)]{qn99}
    Quataert, E., \&  Narayan, R., 1999, \apj, 520, 298
\bibitem[Rawlings \& Saunders(1991)]{rs91}
  Rawlings S., \& Saunders R., 1991, Nature, 349, 138
\bibitem[Romanova et al.(1998)]{r98}
      Romanova M. M., Ustyugova G. V., Koldoba A. V.,
      Chechetkin V. M., \& Lovelace R. V. E.,  1998, \apj, 500, 703
 \bibitem[Serjeant et al.(1998)]{s98}
   Serjeant, S., Rawlings, S., Lacy, M., Maddox, S. J., Baker, J. C., Clements, D., Lilje,
   P. B., 1998, \mnras, 294, 494
\bibitem[Shakura \& Sunyaev(1973)]{ss73}
      Shakura, N. I., \& Sunyaev, R. A., 1973, \aap, 24, 337
\bibitem[Svensson, \& Zdziarski(1994)]{sz94}
      Svensson, R., \& Zdziarski, A., 1994, \apj, 436, 599
\bibitem[Tout \& Pringle(1996)]{tp96}
      Tout C. A., \& Pringle J. E., 1996, \mnras, 281, 219
 \bibitem[Urry \& Padovani(1995)]{up95}
     Urry, C.M., \& Padovani, P., 1995, \pasp, 107, 803
\bibitem[Voges et al.(1999)]{v99}
   Voges, W. et al., 1999, \aap, 349, 389
\bibitem[Willott et al.(1999)]{w99}
    Willott, C.J., Rawlings S., Blundell K.M., \& Lacy M., 1999, \mnras. 309. 1017
\bibitem[Wills \& Browne(1986)]{wb86}
    Wills, B.J., \& Browne, I.W.A., 1986, \apj, 302, 56
\bibitem[Xu, Livio, \& Baum(1999)]{x99}
      Xu, C., Livio, M., Baum, S., 1999, \aj, 1999, \aj, 118, 1169



\end{thebibliography}
\end{document}